\documentclass[twocolumn,superscriptaddress,showpacs,aps,amsmath,amssymb]{revtex4}
\usepackage{graphicx}
\usepackage{amsmath}
\usepackage{amsxtra}
\usepackage{color}

\begin{document}

\title{General non-Markovian dynamics of open quantum systems}

\author{Wei-Min Zhang}
\email{wzhang@mail.ncku.edu.tw}
\affiliation{Department of Physics, National Cheng Kung
University, Tainan 70101, Taiwan}
\affiliation{Advanced Science Institute, RIKEN, Saitama 351-0198, Japan}
\author{Ping-Yuan Lo}
\affiliation{Department of Physics, National Cheng Kung University, Tainan 70101,
Taiwan}
\author{Heng-Na Xiong}
\affiliation{Department of Physics, National Cheng Kung University, Tainan 70101,
Taiwan}
\author{Matisse Wei-Yuan Tu}
\affiliation{Department of Physics, National Cheng Kung University, Tainan 70101,
Taiwan}
\affiliation{Advanced Science Institute, RIKEN, Saitama 351-0198, Japan}
\author{Franco Nori} 
\email{fnori@riken.jp}
\affiliation{Advanced Science Institute, RIKEN, Saitama 351-0198, Japan}
\affiliation{Physics Department, The University of Michigan, Ann Arbor, Michigan, 48109-1040, USA}

\begin{abstract}

We present a general theory of non-Markovian dynamics  for open quantum 
systems.  We explore the non-Markovian dynamics by connecting 
the exact master equations with the non-equilibirum Green 
functions. Environmental back-actions  are fully taken into account.  The
non-Markovian dynamics consists of non-exponential decays and 
dissipationless oscillations. Non-exponential decays are induced by the
discontinuity in the imaginary part of the self-energy corrections. 
Dissipationless oscillations arise from band gaps or the finite band structure of spectral densities.
The exact analytic solutions for various non-Markovian environments show 
that the non-Markovian dynamics can be largely understood from the  
environmental-modified spectra of the open systems.

\end{abstract}


%\keywords{ }

\maketitle

Understanding the dynamics of open quantum systems is one of the
most challenging topics in physics,
chemistry, and biology. The environment-induced quantum dissipation 
and decoherence dynamics are the main concerns in the study of 
open quantum systems \cite{Bre02,Wei08}.  Decoherence control has also recently become 
a key task for practical implementations of nanoscale solid-state quantum 
information processing \cite{Iul09108}, where the decoherence
is mainly dominated by non-Markovian dynamics due to the strong back-actions 
from the environment.  A fundamental issue is 
how to accurately take into account non-Markovian memory effects,
which have attracted considerable attention very recently both in 
theory \cite{An07,Paz08,Wolf08, Tu08235311,Bre09, Koss10, 
Xio10012105,Ang10, Mar11} and in
experiments \cite{Mad11,Liu11,Tang12}.

The non-Markovian dynamics of an open quantum system can be described by 
the master equation of the reduced density matrix $\rho(t)$. This is obtained by tracing 
over the environmental degrees of freedom, $\rho(t) =
\text{tr}[\rho_{\text{tot}}(t)]$, where $\rho_{\text{tot}}(t)$ is
the density matrix of the total system.  The standard approach to 
the non-Markovian dynamics uses 
the Nakajima-Zwanzig operator projective technique \cite{N58Z60}
where the master equation is formally written  as
\begin{align}
\frac{d\rho(t)}{dt}=\! \! \int_{t_0}^t \!\!\! d\tau{\cal K}(t-\tau)\rho(\tau) . \label{nzme}
\end{align}
The non-Markovian
memory effects are taken into account by the time non-local integral kernel 
${\cal K}(t-\tau)$. In practice, very few systems can be exactly solved
from (\ref{nzme}). Therefore, the generality of non-Markovian 
dynamics has not been fully understood.

In general, there are three typical timescales in an open system to
characterize non-Markovian dynamics: (i) the timescale of
the system $\sim 1/\varepsilon_s$, where $\varepsilon_s$ 
is a typical energy scale of that system; (ii) the timescale of
the environment $\sim 1/d$, where $d$ is the bandwidth of
the environmental spectral density; (iii) the mutual timescale arising
from the coupling between the system and the environment $\sim
1/\Gamma$, where $\Gamma$ is the dominant coupling strength.
%and (iv) the thermal and chemical timescales,
%$\sim 1/k_{\rm B}T$ and $1/\mu$, where $T$ and $\mu$ are the
%temperature and the chemical potential of the environment. 
%The back-actions become strong when the characteristic frequency 
%of the system is in resonance with the noise spectrum of  the environment.
%namely the characteristic frequency of the system is centered at 
%the noise spectrum of the environment. 
It is usually believed that non-Markovian memory effects strongly rely
on the relations among these different timescales. However, such relationships 
have not been quantitatively established yet. 

Here, we show that the
general behavior of non-Markovain dynamics is mainly determined
by the band structure of the environment and the coupling
between the system and the environment.
%and a low environmental temperature also plays an important role. 
We explore the non-Markovian dynamics from 
the analytical solution, solved by connecting the exact master equation 
with the non-equilibirum Green functions. Exact master equations 
have been derived only for limited system-environment couplings 
\cite{Paz92,Tu08235311, Jin10083013, Lei121408}. Establishing 
the connection between the master equation and the non-equilibrium  
Green functions provides a new way to explore the non-Markovian dynamics 
even if the exact master equation of the open system is unknown.

\textit{Exact master equation and non-equilibrium Green functions.---}We 
begin with a fermonic (bosonic) many-body system consisting of $N$ 
single-particle energy levels $\varepsilon_i~ (i=1, 2, \cdots, N)$,  coupled, 
via particle-particle exchanges, to a  fermonic (bosonic) environment,
$H_{SB}=\sum_{\alpha k i}[V_{\alpha k i}a^\dag_i b_{\alpha k} + 
V^*_{\alpha k i}b^\dag_{\alpha k}a_i]$.  The environment can 
contain many different reservoirs,  each reservoir is specified  by its spectral density
$\boldsymbol{J}_{\alpha ij}(\omega)=2\pi \sum_{k} V_{\alpha k i}
V^*_{\alpha k j}\delta(\omega \! - \! \epsilon_k)$, where
$V_{\alpha k i}$  is a coupling strength between the 
system and reservoir $\alpha$.  The operators $a^\dag_i$ ($a_i$) and $b^\dag_{\alpha k}$
($b_{\alpha k}$) are the particle creation (annihilation) operators of the dicrete 
energy level $i$ of the system and the continuous level $k$ of reservoir $\alpha$,
respectively. These creation-annihilation operators obey the standard anticommutation 
(commutation) relationship for fermions (bosons).
Nonlinear particle-particle interactions in the system are not included. 
Using the coherent-state path-integral method
\cite{Zhang90} to the Feynman-Vernon influence functional  
\cite{Fey63118}, the exact master equation of such an open system
can be derived \cite{Tu08235311,Jin10083013,Lei121408}:
%\begin{subequations}
\begin{align}
\frac{d\rho(t)}{dt} = & \frac{1}{i}\Big[\widetilde{H}_S(t),
\rho(t)\Big]  + \sum_{ij} \Big\{\gamma_{ij}(t)
\Big[2a_j\rho(t)a^{\dag}_i \nonumber \\
& - a^{\dag}_ia_j\rho(t) -\rho(t) a^{\dag}_ia_j\Big] +
\widetilde{\boldsymbol{\gamma}}_{ij}(t)\Big[a^\dag_i \rho(t)
a_j \nonumber \\ & \pm a_j\rho(t)a^{\dag}_i - a^{\dag}_ia_j\rho(t) 
\mp \rho(t) a_ja^{\dag}_i \Big] \Big\}\, . \label{eme0}
\end{align}
%\end{subequations}
The first term in (\ref{eme0}) is the unitary term with the 
renormalized Hamiltonian $\tilde{H}_S(t)=\sum_{ij}
\widetilde{\boldsymbol{\varepsilon}}_{s ij}(t)a^\dag_i a_j$. 
The second and third terms give the non-unitary dissipation and fluctuations,
respectively.  The $\pm$ and $\mp$ signs in the third term correspond to the
system being bosonic/fermionic. The renormalized energy levels 
$\widetilde{\boldsymbol{\varepsilon}}_s(t)$,  
the time-dependent dissipation coefficient $\boldsymbol{\gamma}(t) $ and 
the fluctuation coefficient $\widetilde{\boldsymbol{\gamma}}(t)$  
in (\ref{eme0}) 
are given by
\begin{subequations}
\label{coeffs}
\begin{align}
& ~~\widetilde{\boldsymbol{\varepsilon}}_s
(t)=\frac{i}{2}\big[\dot{\boldsymbol{u}}(t,t_0)\boldsymbol{u}^{-1}
(t,t_0) -\text{H.c.} \big],   \label{ren-en}
\\ &\boldsymbol{\gamma}(t) = -\frac{1}{2}\big[
\dot{\boldsymbol{u}}(t,t_0)\boldsymbol{u}^{-1}(t,t_0) +
\text{H.c.}\big] , \label{diss-coe} \\
\widetilde{\boldsymbol{\gamma}}(t) &=
\dot{\boldsymbol{v}}(t,t) \!-\! \big[ \dot{\boldsymbol{u}}(t,t_0)
\boldsymbol{u}^{-1}(t,t_0)\boldsymbol{v}(t,t) + \text{H.c.}\big] .  \label{fluc-coe}
\end{align}
\end{subequations}

In Eqs.~(\ref{coeffs}),  the $N\times N$ matrix functions $\boldsymbol{u}(t,t_0)$ and 
$\boldsymbol{v}(t,t)$ are related to the non-equilibrium Green 
functions of the system
%, introduced originally by Schwinger  \cite{Sch61407},  Kadanoff and Baym 
%\cite{Kad62}, and also Keldysh \cite{Kel651018} 
in the Schwinger-Keldysh non-equilibrium theory \cite{Sch61407, Kad62}, 
$\boldsymbol{u}_{ij}(t,t_0)\!=\! \langle [a_i(t), a^\dag_j(t_0)]_\mp \rangle$,
and $\boldsymbol{v}_{ij}(t,t) = \langle a^\dag_j(t) a_i(t)\rangle$ subtracting an 
initial-state dependent part \cite{explain}.  
These Green functions obey the Dyson equations,
\begin{subequations}
\label{kde}
\begin{align}
\frac{d}{d\tau}\boldsymbol{u}(\tau,t_0) + i \boldsymbol{\varepsilon}_s
\boldsymbol{u}(\tau,t_0) \, + & \! \int_{t_0}^{\tau}\!\!\! d\tau' \boldsymbol{g}(\tau,\tau')
\boldsymbol{u}(\tau',t_0) = 0 , \label{rgde}  \\
\frac{d}{d\tau}\boldsymbol{v}(\tau,t) + i \boldsymbol{\varepsilon}_s
\boldsymbol{v}(\tau,t) \, + & \! \int_{t_0}^{\tau} \!\!\! d\tau' \boldsymbol{g}(\tau,\tau')
\boldsymbol{v}(\tau',t)  \nonumber \\
 =& \int^t_{t_0}\!\!\! d\tau' \widetilde{\boldsymbol{g}}(\tau,
\tau') \boldsymbol{u}^\dag(\tau',t_0) ,  \label{cgde}
\end{align}
\end{subequations}
subjected to the boundary conditions $\boldsymbol{u}(t_0,t_0) = 1$
and $\boldsymbol{v}(t_0,t)=0$ with $ t_0 \le \tau \le t$, where $ \boldsymbol{\varepsilon}_s$ 
is a $N\times N$ matrix given by the bare single-particle energy levels of the system.
%, and $\bar{\boldsymbol{u}}(\tau)= \boldsymbol{u}^\dag(t+t_0-\tau)$ is the advanced Green function. 
%The inhomogenous term in the r.h.s of (\ref{cgde}) counts the fluctuations from the environment.
%and the nonequilibrium thermal fluctuation is characterized by
%the correlation Green function $\boldsymbol{v}(t, \tau)$ which is given by
The self-energy corrections, $\boldsymbol{g}(\tau,\tau')$ and 
$\widetilde{\boldsymbol{g}}(\tau, \tau')$, which take into account all the 
back-actions from the environment, are expressed explicitly by
\begin{subequations}
\label{self-en}
\begin{align}
\label{timecorrel-1} & \boldsymbol{g}(\tau,\tau') = \sum_\alpha
\int \frac{d\omega}{2\pi}\boldsymbol{J}_\alpha(\omega)e^{-i\omega(\tau-\tau')} \ , \\
\label{timecorrel-2} & \widetilde{\boldsymbol{g}}(\tau,\tau') = \sum_\alpha
\int\frac{d\omega}{2\pi}\boldsymbol{J}_\alpha(\omega)f_\alpha(\omega)
e^{-i\omega(\tau-\tau')} \ ,
\end{align}
\end{subequations}
where the function $f_\alpha(\omega) =
[e^{\beta_\alpha(\omega-\mu_\alpha)} \mp 1]^{-1}$ is the Bose-Einstein
(Fermi-Dirac) distribution of bosonic (fermionic)
reservoir $\alpha$ at the initial time $t_0$. Equations (\ref{eme0})-(\ref{self-en}) 
establish a rigorous connection between the known exact master equation 
and the non-equilibrium Green functions for open quantum systems.

\textit{General non-Markovian dynamics.---}Different from the Nakajima-Zwanzig
master equation, the exact master equation (\ref{eme0}) is local in time,
characterized by the dissipation and the fluctuation coefficients, $\boldsymbol{\gamma}(t)$
and $\widetilde{\boldsymbol{\gamma}}(t)$. 
%The appearance of $t_0$ in these coefficients indicates the memory of the initial time. 
Non-Markovian memory effects are manifested as follows: 
 
(i) The coefficients $\boldsymbol{\gamma}(t)$ and $\widetilde{\boldsymbol{\gamma}}(t)$ 
 are microscopically and non-perturbatively determined by the non-equilibrium 
 Green functions from the Dyson equations (\ref{kde}). The non-Markovian 
 memory effect is fully coded into the homogenous non-local 
 time integrals in (\ref{kde}) with the integral kernel ${\boldsymbol{g}}(\tau,\tau')$.
 In other words, the self-energy correction $\boldsymbol{g}(\tau,\tau')$  serves as 
 a memory kernel that count all the back-actions  from the environment. 
 
 (ii) The coefficients $\boldsymbol{\gamma}(t)$ and $\widetilde{\boldsymbol{\gamma}}(t)$ 
 are constrained by the non-equilibrium fluctuation-dissipation theorem. 
The inhomogenous non-local time integral in (\ref{cgde}) with the integral 
kernel $\widetilde{\boldsymbol{g}}(\tau,\tau')$, depicts the fluctuation arisen from the
 environment.  Because $\boldsymbol{v}(t_0,t)=0$, we can 
 analytically solve Eq.~ (\ref{cgde}): 
\begin{align}
\boldsymbol{v}(\tau,t) = \int^\tau_{t_0}\!\!\! d\tau_1 \!\! \int^t_{t_0}\!\!\!
d\tau_2 \, \boldsymbol{u}(\tau, \tau_1)\, \widetilde{\boldsymbol{g}}(\tau_1, \tau_2) \,
\boldsymbol{u}^\dag(t,\tau_2) \ .  \label{nedft}
\end{align} 
This solution shows that Eq.~(\ref{fluc-coe}) is a generalized non-equilibrium 
fluctuation-dissipation theorem in the time domain (the reduction to the equilibrium 
fluctuation-dissipation theorem is given in \cite{supm}).  The fluctuation-dissipation 
theorem is a consequence of the unitarity of the whole system. It guarantees the 
positivity of the reduced density matrix during the non-Markovian time 
evolution. 
%In contrast, in the Lindblad-GKS master equation, these coefficients are mathematically 
%required to be time-independent, and must be positive to ensure that the Lindblad-GKS master 
%equation gives complete positive, trace-preserving maps.
%\iffalse 
%(iii) Besides the time-dependence, by solving the Dyson equation, it shows 
%that $\boldsymbol{\gamma}(t)$ and $\widetilde{\boldsymbol{\gamma}}(t)$  
%can vary sequentially from positive to negative values \cite{Xio10012105}, 
%representing  the back-flow of information from the system to the environment \cite{Liu11}. 
%as a typical signal of  non-Markovian dynamics. 
%In contrast, the time-independent, 
%positive rate coefficients in the Lindblad-GKS master equation give only the simple 
%monotonous exponential decays, as a typical feature of  Markovian dynamics. 
%In the weak-coupling limit, the solution of (\ref{kde}) with a perturbation 
%expansion up to second order in the coupling strength between the 
%system and the environment leads to the Born-Markovian 
%master equation that has been studied widely in the literature  \cite{Bre02,Wei08},
%where $\boldsymbol{\gamma}(\tau)$ and $\widetilde{\boldsymbol{\gamma}}(t)$ 
%become positive (see the supplementary materials \cite{supm}, and also 
%Ref.~\cite{Tu08235311, Xio10012105}).
%\fi

Based on the above intrinsic features of open quantum systems, we can now 
explore the general properties of non-Markovian dynamics. From Eqs.~(\ref{coeffs}), 
we can  express the Green function $\boldsymbol{u}(t,t_0)$ in terms of  
the dissipation coefficient $\boldsymbol{\gamma}(t)$ as
\begin{align}
\boldsymbol{u}(t,t_0) = {\cal T}  \exp \Big\{\! \! -\!\! \int^t_{t_0}\!\!\! 
d\tau \big[i \widetilde{\boldsymbol{\varepsilon}}(\tau)+\boldsymbol{\gamma}(\tau) \big] \Big\},
\end{align}
where ${\cal T}$ is the time-ordering operator. This solution indicates
that $\boldsymbol{u}(t,t_0)$ fully determines the dissipation
dynamics of the system.  However, due to the time-dependence of the 
dissipation coefficients,  the detailed dissipation dynamics can vary 
significantly for different environments. 

\begin{figure}
\setlength{\unitlength}{1mm}
\begin{picture}(70,35)(0,0)
\linethickness{0.3mm}
  \put(0,15){\vector(1,0){72}}  
  \put(67,11){${\rm Re}[z]$}
  \put(0,7){\vector(1,1){15}} 
  \put(12.5,23){${\rm Im}[z]$}
  \put( 8,5){\vector(0,1){22}} 
  \put( 0,26){$U(z)$}
  \linethickness{0.8mm}
  %\color{red}
  \put(16,15){\color{red}\line(1,0){10}} 
%  \put(16,13){\color{red}\line(0,1){4}}   
%  \put(16,13){\color{red}\line(1,1){4}}  
%  \Photon(16,15)(26,15){4}{4.5}  
  \put(45,15){\color{red}\line(1,0){25}}    
% \dottedline{3}(1,0){16}}  
  \linethickness{0.2mm}
%  \multiput(13,9)(20,0){2}{\line(0,0){5}}   
  \put(11,5){$\omega_1$}
  \put(12,8){\vector(0,1){5}}
  \put(30,5){$\omega_2$}
  \put(31,8){\vector(0,1){5}}
  \put(39,5){$\omega_3$}
  \put(40,8){\vector(0,1){5}}
  \multiput(12,15)(19,0){2}{\circle*{2}}
  \put(40,15){\circle*{2}}
%  \put(40,8){\vector(0,1){5}}  
  \put(20,17){$B_1$}
%  \put(28,22){\vector(-1,1){14}} 
  \put(48,17){$B_2$}
%  \put(52,17){$B_3$}
 % \multiput(0,37)(4,0){13}{\line(1,0){2}}
\end{picture}
\caption{(color online) A schematic pole structure of the Green function $\boldsymbol{U}(z)$. The thick red lines 
on the real $z$ axis correspond to $\sum_\alpha \boldsymbol{J}_\alpha(z)\ne 0$.}
\label{fig1}
\end{figure}
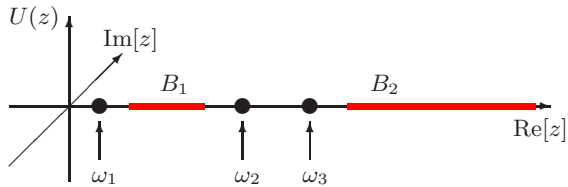

Explicitly, equation (\ref{self-en}) show that $\boldsymbol{g}(\tau,\tau')=\boldsymbol{g}(\tau\!-\!\tau')$
and $\widetilde{\boldsymbol{g}}(\tau,\tau')=\widetilde{\boldsymbol{g}}(\tau\!-\!\tau')$. Thus
we can write  $\boldsymbol{u}(t,t_0)=\boldsymbol{u}(t-t_0)$. Using the modified Laplace 
transform $\boldsymbol{U}(z)=\!\int^\infty_{t_0}\!\!dt \,\boldsymbol{u}(t)e^{iz(t-t_0)}$, it is easy to
obtain
\begin{align}
\boldsymbol{U}(z) = \frac{i}{z{\rm \bf{I}} - \boldsymbol{\varepsilon}_s 
-\boldsymbol{\Sigma}(z)} ,
\label{delt}
\end{align}
where {\textbf{I}} is the identity,  $\boldsymbol{\Sigma}(z)$ is the 
Laplace transform of the self-energy correction,
\begin{align}
\boldsymbol{\Sigma}(z)=\sum_\alpha\int \frac{d\omega}{2\pi} 
\frac{\boldsymbol{J}_\alpha(\omega)}{z - \omega}
\overset{z=\omega \pm i0^+ }{\longrightarrow} \boldsymbol{\Delta}(\omega) 
\mp i \sum_\alpha\frac{\boldsymbol{J}_\alpha(\omega)}{2} ,  \label{ltse}
\end{align}
and $\boldsymbol{\Delta}(\omega)= \sum_\alpha{\cal P}\int 
\frac{d\omega'}{2\pi} \frac{\boldsymbol{J}_\alpha(\omega')}{\omega - \omega'} $ 
is the principal value of the integral. It can be shown that
the general solution of $\boldsymbol{u}(t,t_0)$ is given by
\begin{align}
\boldsymbol{u}(t  - t_0) =  \sum_i \boldsymbol{\cal Z}_i & e^{-i\omega_i (t-t_0)} +
\sum_k \int_{B_k} \frac{d\omega}{2\pi} \Big[\boldsymbol{U}(\omega+i0^+)
 \nonumber \\
&~~~~~~ -\boldsymbol{U}(\omega-i0^+) \Big] 
e^{-i\omega(t-t_0)} . \label{u-gs}
\end{align}
%where the first term comes from the poles $\{\varepsilon_i\}$ of Eq.~(\ref{delt}) 
%and $\{\boldsymbol{\cal Z}_i \}$ are the corresponding residues, 
%, and $\boldsymbol{u}^{\Pi}_k(s)$ and $\boldsymbol{u}^{\rm I}_k(s)$ 
%are defined on the first and the second Riemannian sheets. 
%\begin{figure}[htbp] 
%\centering \includegraphics[width=2in]{retardedgf.pdf} 
%\caption{The general structure of the retarded Green function}
%\end{figure}
The first term in (\ref{u-gs}) corresponds to localized modes with poles
$\{\omega_i\}$ located at the real $z$ axis with $\sum_\alpha 
\boldsymbol{J}_\alpha(\omega)=0$. The coefficients 
$\{\boldsymbol{\cal Z}_i \}$ are the corresponding residues.  The localized modes 
exist only when the environmental spectral density has band gaps or a finite band, i.e., 
$\sum_\alpha \boldsymbol{J}_\alpha(\omega)$ vanishes 
in some frequency regions, see Fig.~\ref{fig1}.  These localized modes 
do not decay, and give dissipationless non-Markovian dynamics. 
The second term in (\ref{u-gs}) is the contribution from the branch cuts $\{B_k \}$, 
due to the discontinuity of $\boldsymbol{\Sigma}(z)$, so does $\boldsymbol{U}(z)$, 
across the real axis on the complex space
$z$, see Eq.~(\ref{ltse}). The branch cuts usually generate non-exponential decays 
\cite{Tannoudji92}, which is another significance of the non-Markovian dynamics. 
When the system is weakly coupled to the environment, the non-exponential 
decays are reduced to exponential-like decays.

Equation (\ref{u-gs}) provides indeed a general solution of 
the non-Markovian dissipation dynamics. It shows that the non-Markovian 
dissipation dynamics consists of  non-exponential 
decays plus dissipationless localized modes. 
%This is the general dynamic feature of any microscopic 
%particle (considered as the principal system) moving in 
%many-body systems (treated as the environment). 
Such a solution for the two-point Green function $\boldsymbol{u}(t,t_0)$
is generic and can be proven from the quantum field theory \cite{Peskin95},  
even if particle-particle interactions are included. 

The Green function $\boldsymbol{u}(t,t_0)$
reveals the general non-Markovian dissipation dynamics. 
The non-Markovian fluctuation dynamics is constrained by
the fluctuation-dissipation theorem via the Green function
$\boldsymbol{v}(t,t)$ of (\ref{nedft}).
%\iffalse 
%The environmental fluctuations can also generate complicated non-Markovian 
%dynamics if the environment is initially at low temperatures.  In fact, 
%even for a constant spectral density,
%$\boldsymbol{J}_\alpha(\epsilon)=\boldsymbol{\Gamma}_\alpha$ (wide-band limit),
%the dissipation memory kernel is reduced to a delta function 
%$\boldsymbol{g}(\tau, \tau')= \delta(\tau-\tau')\sum_\alpha
%\boldsymbol{\Gamma}_\alpha/2$. 
%However, the fluctuation memory kernel,   
%\begin{subequations}
%\begin{align}
%\widetilde{\boldsymbol{g}}(\tau,\tau') = &\mp \sum_\alpha \Big[
%\frac{\boldsymbol{\Gamma}_\alpha}{2}
%\delta(\tau-\tau')   \mp \frac{\boldsymbol{\Gamma}_\alpha}{\beta_\alpha} \sum_{m=0}^\infty
%e^{-\gamma_{\alpha m} (\tau-\tau')} \Big], \nonumber %\label{gambeta2}
%\end{align}
%\end{subequations}
%where $\gamma_{\alpha m}=2m\pi /\beta_\alpha ~[ (2m+1)\pi/\beta_\alpha-i\mu_\alpha]$ 
%is the Bose [Fermi] Matsubara frequency, remains {\it non-local} in time \cite{Jin10083013}. 
%This may give  non-Markovian fluctuation effects. 
%Only at high temperatures, the summation term in Eq.~(\ref{gambeta2}) 
%becomes a delta function of $(\tau-\tau')$ so that the truly Markovian limit is reached.
%\fi
Thus, the whole picture of non-Markovian dynamics is fully characterized  
by the dissipation and fluctuation coefficients of (\ref{coeffs}). The non-exponential 
decay part of (\ref{u-gs}) makes the dissipation coefficient $\boldsymbol{\gamma}(t)$  
oscillates between positive and negative values, representing the back-flow of information 
from the system to the environment \cite{Bre09,Koss10}. Non-exponential 
decays alone give $\boldsymbol{\gamma}(t)$ a non-zero asymptotical value. If there are 
localized modes, $\boldsymbol{\gamma}(t)$ will vanish in the steady-state limit, 
resulting in dissipationless oscillations.
In the weak coupling region, $\boldsymbol{\gamma}(t)$ can still be time-dependent
but keeps positive, the corresponding dynamics gives simple exponential-like decays,
observed mainly in the Markovian limit.  
%With the help of the exact master equation, we can find that 
%\begin{align}
%\boldsymbol{v}_{ij}(t,t)  \! = \!  \langle a^\dag_j(t)a_i(t) \rangle\!-\! \boldsymbol{u}_{ii'}(t,t_0) 
%\langle a^\dag_{j'}(t_0)a_{i'}(t_0) \rangle \boldsymbol{u}^\dag_{j'j}(t,t_0) 
% . \label{ccf}
%\end{align}
%This shows that $\boldsymbol{v}(t,t)$ is 
%the environment-induced particle density matrix of the system.  
Furthermore Eqs.~(\ref{fluc-coe}) and (\ref{nedft}) together show that 
except for the initial environmental temperature dependence, the 
time-dependence of the fluctuation coefficient  $\boldsymbol{\widetilde{\gamma}}(t)$ 
behaves similar to $\boldsymbol{\gamma}(t)$, due to the fluctuation-dissipation 
theorem. In conclusion, non-Markovian dynamics can be fully understood 
from the solution of the Green function $\boldsymbol{u}(t,t_0)$.  

\begin{figure}[h]
\includegraphics[width=0.48\textwidth]{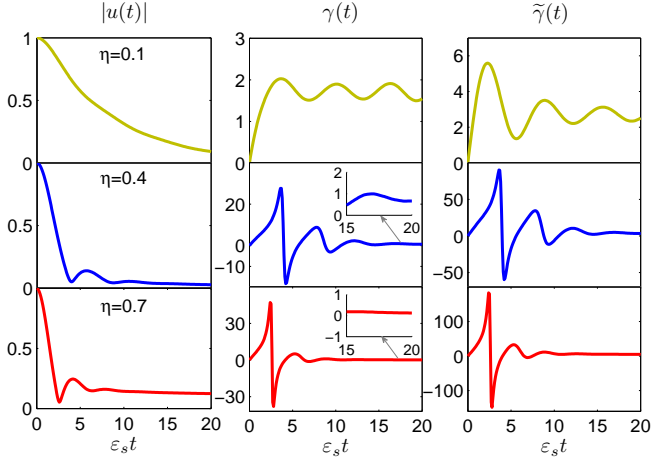}
\caption{(color online) The time evolution of the Green function $u(t)$, the dissipation
and the fluctuation coefficients, $\gamma(t)$ and $\widetilde{\gamma}(t)$, in a 
sub-Ohmic bath, for several different values of the coupling constant $\eta$. We take other parameters
$\varepsilon_s=13.83 \mu$eV, $\omega_c=\varepsilon_s$ and also $k_BT=\varepsilon_s$.}
\label{fig2}
\end{figure}

\textit{Examples and discussion ---}To be more specific,  let us first examine the 
non-Markovian dynamics of a single-mode bosonic nanosystem, such as 
a nanophotonic or optomechanical resonator, coupled to
a general non-Markovian environment with spectral density
\begin{align}
J(\omega)=2\pi\eta \omega \Big(\frac{\omega}{\omega_c}\Big)^{s-1} 
\!\!\!\!\exp\Big(\! \! -\!  \frac{\omega}{\omega_c}\Big),  %\nonumber
\end{align}
where $\eta$ is  the coupling constant 
between the system and the environment,  and $\omega_c$ is the frequency cutoff. 
When $s=1$, $<1$ and $>1$, the corresponding environments are Ohmic, sub-Ohmic 
and super-Ohmic, respectively \cite{Leg871}.  Following the above general 
procedure, the analytical solution of the non-Markovian dissipation dynamics 
is given by (setting $t_0=0$ for simplicity):
%\begin{subequations}
\begin{align}
u(t)= {\cal Z} e^{-i\omega' t} +\frac{2}{\pi} \int_0^\infty \!\!\! d\omega
 \frac{J(\omega)e^{-i\omega t}}{4[\omega-\varepsilon_s 
 -\Delta(\omega)]^2 + J^2(\omega)} \, ,  %\nonumber
\end{align}
where  $\Delta(\omega)=\frac{1}{2}[\Sigma(\omega+i0^+)
+ \Sigma(\omega-i0^+)]$ and the Laplace transform of the self-energy correction 
\begin{align}
\Sigma(\omega)= \left\{ \begin{array}{ll} \eta \omega_c 
\big[\pi\sqrt{-\widetilde{\omega}} e^{-\widetilde{\omega}}
{\rm erfc}(\sqrt{-\widetilde{\omega}})-\sqrt{\pi}\big] & s=1/2 \\ & \\
\eta \omega_c \big[\widetilde{\omega} \exp(-\widetilde{\omega})
{\rm Ei}(\widetilde{\omega})-1\big] & s=1\\ & \\
\eta \omega_c \big[\widetilde{\omega}^3 e^{-\widetilde{\omega}}
{\rm Ei}(\widetilde{\omega}) - \widetilde{\omega}^2
- \widetilde{\omega}-2\big] & s=3 \end{array} \right.  %\nonumber
\end{align}
%\end{subequations}
with $\widetilde{\omega}=\omega/\omega_c$. 
Due to the vanishing spectral density for $\omega <0$, a 
localized mode at $\omega'=\varepsilon_s - \Sigma(\omega') <0 $ occurs
when $\eta \omega_c \Gamma(s) > \varepsilon_s$, here $\Gamma(s)$ is a 
gamma function. The localized mode leads to the dissipationless process. 
The corresponding  residue is  ${\cal Z}=[1-\Sigma'(\omega')]^{-1}$.
This analytical solution precisely reproduces the exact numerical solution 
in the previous work \cite{Xio10012105}. Figure \ref{fig2} shows that for 
a small $\eta$, the dissipation dynamics is an exponential-like decay,
The corresponding $\gamma(t)$ and $\widetilde{\gamma}(t)$ are time-dependent 
but positive (corresponding to Markovian dynamics).
When $\eta \gtrsim 0.3$, the non-exponential decay dominates, and $\gamma(t)$ and 
$\widetilde{\gamma}(t)$ oscillate in positive and negative values with nonzero asymptotical 
values. When $\eta \gtrsim 0.6$, the localized state occurs, and $u(t)$ does not decay to
zero. Correspondingly, $\gamma(t)$ and $\widetilde{\gamma}(t)$ asymptotically 
approach to zero.

The second example is a fermonic system, a single electron transistor 
in nanostructures which consists of 
a quantum dot coupled to a source and a drain. The source and the drain are 
treated as two reservoirs of the environment. Their spectral densities take a 
Lorentzian form with a sharp cutoff,
\begin{align}
J_\alpha(\omega)= \frac{\Gamma_\alpha d^2_\alpha}{(\omega - 
\omega_c) ^{2} + d^{2}_\alpha } 
\Theta ( \Omega - | \omega - \omega_c|)   %\nonumber
\end{align}
with $\alpha=L (R)$ for the source (drain), where $d_\alpha$ is the halfwidth of the spectral density
and  $\Gamma_\alpha$ is the coupling strength between the system and reservoir $\alpha$.
We add a sharp cutoff to simulate a finite band for the environmental density of states. 
When $\Omega \rightarrow \infty$,  the above spectral density is reduced to 
the usual Lorentzian spectral density that has been used  in various studies of 
nanoelectronics \cite{Mei93, Bla00,Wal06, Jin08,Tu08235311}. 
We consider the symmetric case, $(\Gamma_L, d_L)=(\Gamma_R,d_R)=(\Gamma,d)$.
Then the analytical solution of the Green function $u(t)$ becomes
%\begin{subequations}
\begin{align}
u(t) = &\sum^2_{j=1} {\cal Z} _{j} e ^{- i \omega'_j t } 
+ \frac{1}{\pi} \int_{\omega_c- \Omega} ^{\omega_c+ 
\Omega} \!\!\! \!\! \!\!\!d \omega \frac{J( \omega ) 
e ^{- i \omega t }}{[ \omega - \varepsilon_s - 
\Delta ( \omega )] ^{2} + J^{2} ( \omega ) },  %\nonumber
\end{align}
where $J(\omega)=J_L(\omega)=J_R(\omega)$ and $\Delta(\omega)$ is the 
real part of the self-energy $\Sigma(\omega)$,
\begin{align}
\Sigma (\omega)= \frac{J(\omega)}{\pi} \Bigg[\log \frac{\omega_c
-\Omega -\omega}{\omega_c+\Omega - \omega}
 +   \frac{2 (\omega-\omega_c) }{d}\tan ^{-1} \frac{\Omega} 
 {d}\Bigg].  
\end{align}
%\end{subequations}
The two localized states are located outside of the band, i.e.,
$\omega'_j=\varepsilon_s + \Sigma(\omega'_j)$, with $\omega'_1 < \omega_c -\Omega$, and 
$\omega'_2 >\omega_c + \Omega$. The corresponding residue is given by
${\cal Z}_j=[1-\Sigma'(\omega'_j)]^{-1}$. Again, the localized modes lead to 
a dissipationless process and the integral term shows a non-exponential 
decay. Taking $\Omega \rightarrow \infty$,  
the two localized modes are excluded, and the solution of $u(t)$ reproduces
the exact non-Markovian dynamics of the usual Lorentzian spectral density 
(for detailed derivation, see \cite{supm}). 

The third example is a two-level system involving single-photon processes
(spontaneous emission). In general, a multi-level atomic open system does not
obey the master equation (\ref{eme0}). However,
the Schr\"{o}dinger equation of a two-level atomic system 
with only spontaneous single-photon emission processes (at
zero temperature) can be reduced to the Dyson equation of (\ref{rgde}) 
\cite{John94,expn1, Gar972290}. For a two-level artificial atom, such as a quantum dot,  
embedded in photonic crystals, because of the photonic band gap
it was shown  \cite{expn1} that the corresponding solution contains exponential 
decays, non-exponential decays, and localized bound modes all together. We find 
analytically \cite{Lo12} that the complex pole with exponential decay shown in 
\cite{expn1} has been included in the brach-cut integral of (\ref{u-gs}). 
Explicitly, the spectral density of the photonic crystals takes 
$J(\omega)=\frac{2C}{\sqrt{\omega-\omega_e}}\Theta(\omega-\omega_e)$ 
\cite{John94,expn1}. From Eq.~(\ref{u-gs}), we directly obtain the 
analytical solution of the spontaneous emission dynamics   
\begin{align}
u(t)=\frac{2\omega_r}{3\omega_r\!+\!\Delta}e^{i(\omega_r-\omega_e) t} +\frac{C}{\pi}
\!\! \int_{\omega_e}^\infty \!\!\! \!\! \! d\omega \frac{\sqrt{\omega\!-\!\omega_e}e^{-i\omega t}}
{(\omega\!-\!\varepsilon_s)^2(\omega\!-\!\omega_e)\!+\!C^2} , 
\end{align}
where $\omega_r$ is the real root given by $(\omega_r+\Delta)\sqrt{\omega_r}=C$, and 
$\Delta=\varepsilon_s-\omega_e$ is the detuning. This analytical solution 
recovers both the exact analytical and numerical solutions given in \cite{John94,expn1}. 

The above examples show that very different open systems coupled to very 
different environments obey the same solution, Eq.~(\ref{u-gs}), of the non-Markovian
dynamics. The solutions of these examples can further be 
written in general as $u(t-t_0)=\int_{-\infty}^\infty \frac{d\omega}{2\pi} {\cal D}(\omega) 
\exp\{-i\omega (t-t_0)\}$ with
\begin{align}
{\cal D}(\omega) = 2\pi\!\! \sum_j {\cal Z}_j \delta(\omega-\omega'_j) 
+ \frac{J(\omega)}{[\omega\!-\!\varepsilon_s \!-\!\Delta(\omega)]^2 + J^2(\omega)/4} .
   \label{spect-d} 
\end{align}
Equation (\ref{spect-d}) shows that the environment modifies the system spectrum 
as a combination of localized modes (dissipationless process) plus a 
continuum spectrum part (non-exponential decays). Remarkably, the result obtained 
from these simple examples gives indeed the general structure of two-point
correlation functions in arbitrary complicated systems, see Ref.~\cite{Peskin95}.
This indicates that alternatively, non-Markovian dynamics can
be fully characterized by the environmental-modified spectrum of the system. If the 
spectrum of the open system can be measured, the non-Markovian dynamics can be 
extracted from its Fourier transform. This largely simplifies the exploration of the
general properties of non-Makovian dynamics for more complicated open systems.

\textit{Conclusion.---}By connecting the exact master equation 
with the non-equilibrium Green functions in many-body systems, 
we derive a general analytical solution of non-Markovian dynamics 
for open quantum systems, i.e., Eq.~(\ref{u-gs}) or (\ref{spect-d}). 
From the analytical solution, we show that the 
general non-Markovian dynamics consists of non-exponential decays 
and dissipationless oscillations. The dissipationless processes arise 
from band gaps or finite band structures of the environmental spectral densities. The 
non-exponential decays are induced by the discotinuity in the imaginary part of the 
self-energy corrections from the environment. The exponential decays 
observed in Markovian limit is a special case in the weak coupling 
limit. Since the non-equilibrium Green 
functions are well-defined for arbitrary quantum systems,  this theory 
may also provide a new approach to explore non-Markovian dynamics for 
more complicated open systems whose exact master equation may be
unknown.

%\begin{acknowledgments}
This work is partially supported by the National Science Council of Republic
of China under Contract No.~NSC-96-2112-M-006-011-MY3. FN is
partially supported by the ARO, NSF grant No.~0726909,
JSPS-RFBR contract No.~12-02-92100, Grant-in-Aid for Scientific Research (S),
MEXT Kakenhi on Quantum Cybernetics, and the JSPS via its FIRST program.

%\end{acknowledgments}


\begin{thebibliography}{99}
\bibitem{Bre02} H. P. Breuer and F. Petruccione, \textit{The Theory of Open
Quantum Systems} (Oxford University Press, New York, 2002).

\bibitem{Wei08} U. Weiss, \textit{Quantum Dissipative Systems} (3rd Ed. World Scientific,
Singapore, 2008).

\bibitem{Iul09108} I. Buluta, S. Ashhab, and F. Nori, Rep. Prog. Phys. \textbf{74}, 104401 (2011).

\bibitem{An07} J. H. An and W. M. Zhang, Phys. Rev. A \textbf{76}, 042127 (2007).

\bibitem{Paz08} J. P. Paz and A. J. Roncaglia, Phys. Rev. Lett. \textbf{100}, 220401 (2008); 
Phys. Rev. A \textbf{79}, 032102 (2009).

\bibitem{Wolf08} M. M. Wolf, J. Eisert, T. S. Cubitt, and J. I. Cirac, Phys. Rev. Lett. \textbf{101}, 150402 (2008).

\bibitem{Tu08235311} M. W. Y. Tu and W. M. Zhang, Phys. Rev. B {\bf 78}, 235311
(2008).

\bibitem{Bre09} H.-P. Breuer, E.-M. Laine, and J. Piilo, Phys. Rev. Lett. \textbf{103}, 210401 (2009); E.-M. Laine, J. Piilo, and H.-P. Breuer, Phys. Rev. A \textbf{81}, 062115 (2010)

\bibitem{Koss10} D. Chru\'{s}ci\'{n}ski, and A. Kossakowski, Phys. Rev. Lett. \textbf{104}, 070406 (2010); D. Chru\'{s}ci\'{n}ski, A. Kossakowski, and A. Rivas, Phys. Rev. A \textbf{83}, 052128 (2011).

\bibitem{Xio10012105} H. N. Xiong, W. M. Zhang, M. H. Wu, and X. G. Wang, 
Phys. Rev. A \textbf{82}, 012105 (2010); C. U Lei and W. M. Zhang, Phys. Rev. A \textbf{84}, 052116 (2011).

\bibitem{Ang10} A. Rivas, S. F. Huelga, and M. B. Plenio, Phys. Rev. Lett. \textbf{105}, 050403 (2010).

\bibitem{Mar11} M. Znidaric, C. Pineda, and I. Garcia-Mata, Phys. Rev. Lett. \textbf{107}, 080404 (2011) .

\bibitem{Liu11} B. H. Liu, L. Li, Y. F. Huang, C.-F. Li, G. C. Guo, E. M. Laine, H.-P. Breuer, and J. Piilo, Nature Phys. \textbf{7}, 931 (2011).

\bibitem{Mad11} K. H. Madsen, S. Ates, T. Lund-Hansen, A. L\"{o}ffler, S. Reitzenstein, A. Forchel, and P. Lodahl, Phys. Rev. Lett. \textbf{106}, 233601 (2011).

\bibitem{Tang12} J.-S. Tang, C.-F. Li, Y.-L. Li, X.-B. Zou, G.-C. Guo, H.-P. Breuer, E.-M. Laine, and J. Piilo, EPL \textbf{97}, 10002 (2012).

\bibitem{N58Z60} S. Nakajima, Prog. Theor. Phys. \textbf{20}, 948 (1958);
R. Zwanzig, J. Chem. Phys. \textbf{33}, 1338 (1960).

\bibitem{Paz92} B. L. Hu, J. P. Paz, and Y. H. Zhang, Phys. Rev. D {\bf 45}, 2843 (1992).

\bibitem{Jin10083013} J. S. Jin, M. W. Y. Tu, W. M. Zhang, and Y. J. Yan, New J. Phys. \textbf{12}, 083013 (2010).

\bibitem{Lei121408} C. U Lei and W. M. Zhang, Ann. Phys. \textbf{327}, 1408 (2012).

\bibitem{Zhang90} W. M. Zhang, D. H. Feng, and R. Gilmore, Rev. Mod. Phys. \textbf{62}, 867 (1990).

\bibitem{Fey63118} R. P. Feynman and F. L. Vernon, Ann. Phys. {\bf 24}, 118 (1963).

\bibitem{Sch61407} J. Schwinger, J. Math. Phys. \textbf{2}, 407 (1961); L. V. Keldysh, Sov. Phys. JETP, \textbf{20}, 1018 (1965).

\bibitem{Kad62} L. P. Kadanoff, and G. Baym, \textit{Quantum Statistical Mechanics}
(Benjamin, New York, 1962).

\bibitem{explain} Explicitly, 
$\boldsymbol{v}_{ij}(\tau,t)= \langle a^\dag_j(t)a_i(\tau) \rangle\! - \! \boldsymbol{u}_{ii'}(\tau,t_0) 
\langle a^\dag_{j'}(t_0)a_{i'}(t_0) \rangle \boldsymbol{u}^\dag_{j'j}(t,t_0)$, where  $\langle \, \cdot \, \rangle$ 
denotes the initial state expectation value. In the standard non-equilibrium 
Green function formalism, $G^<_{ij}(\tau,t)\equiv i\langle a^\dag_j(t)a_i(\tau) \rangle$ is the so-called
lesser Green function. On the other hand, $\boldsymbol{u}_{ij}(t,t_0)\!
=\! \langle [a_i(t), a^\dag_j(t_0)]_\mp \rangle$ is often called 
the spectral Green function, see \cite{Kad62}.

\bibitem{supm} see supplementary materials.

\bibitem{Leg871}  A. J. Leggett, S. Chakravarty, A. T. Dorsey, M. P. Fisher, A. Garg, 
and W. Zwerger, Rev. Mod. Phys. {\bf 59}, 1 (1987).

\bibitem{Tannoudji92} C. Cohen-Tannoudji, J. Dupont-Roc, and G. Grynberg,
\textit{Atom-Photon Interactions} (Wiley, New York, 1992).

\bibitem{Mei93} Y. Meir, N. S. Wingreen, and P. A. Lee, Phys. Rev. Lett. \textbf{70}, 2601 (1993).

\bibitem{Bla00}B. Elattari and S. A. Gurvitz, Phys. Rev. A \textbf{62}, 032102 (2000).

\bibitem{Wal06} S. Welack, M. Schreiber, and U. Kleinekathoferb, J. Chem. Phys. \textbf{124}, 044712 (2006).

\bibitem{Jin08} J. Jin, X. Zheng, and Y. J. Yan, J. Chem. Phys. \textbf{128}, 234703 (2008).

\bibitem{Bre991633} H.-P. Breuer, B. Kappler, and F. Petruccione,
Phys. Rev. A \textbf{59}, 1633 (1999)

\bibitem{Peskin95} M. E. Peskin and D. V. Schroeder, \textit{An Introduction 
to Quantum Field Theory} (Addison-Wesley, Reading, 1995), p.214-215.

\bibitem{John94}S. John and J. Wang, Phys. Rev. Lett. \textbf{64}, 2418 (1990); 
S. John and T. Quang, Phys. Rev. A \textbf{50}, 1764 (1994).

\bibitem{expn1} A. G. Kofman, G. Kurizki, and B. Sherman, J. Mod. Opt. \textbf{41},
353 (1994); A. G. Kofman and G. Kurizki, Phys. Rev. A \textbf{54}, R3750 (1996).

\bibitem{Gar972290} B. M. Garraway, Phys. Rev. A \textbf{55}, 2290 (1997).

\bibitem{Lo12} P. Y. Lo, H. N. Xiong and W. M. Zhang, unpublished (2012).

\end{thebibliography}
\end{document}